\documentclass[copyright,creativecommons]{eptcs}
\providecommand{\event}{}
\providecommand{\volume}{}
\providecommand{\anno}{2009}
\providecommand{\firstpage}{1}

\usepackage{epsfig}
\usepackage{amsfonts}
\usepackage{amssymb}
\usepackage{amsmath}
\usepackage{xcolor}

\newtheorem{theorem}{Theorem}[section]
\newtheorem{definition}[theorem]{Definition}
\newtheorem{proposition}[theorem]{Proposition}

\newtheorem{example}[theorem]{Example}

\newtheorem{remark}[theorem]{Remark}
\newtheorem{notation}[theorem]{Notation}

\newcommand{\qed}{\hfill \ensuremath{\Box}}

\newcommand{\CalculusShortName}{\textsc{CWC}}

\newcommand{\SCalculusLongName}{Stochastic Calculus of Wrapped Compartments}

\newcommand{\blank}{\sharp}

\newcommand{\qqop}[1]{\mathrel{\makebox[2em]{$#1$}}}

\newcommand{\agr}{\quad\big|\quad}

\newcommand{\mydots}{\cdot\cdot\cdot}

\newcommand{\AT}{\mathcal{A}}
\newcommand{\AS}{\overline{\mathcal{A}}}
\newcommand{\TT}{\mathcal{T}}
\newcommand{\TS}{\overline{\mathcal{T}}}

\newcommand{\PP}{\mathcal{S}}
\newcommand{\Pat}{S}
\newcommand{\LeftPP}{\mathcal{P}}
\newcommand{\LeftPat}{P}

\newcommand{\OT}{\mathcal{O}}

\newcommand{\RightPat}{O}

\newcommand{\VarOf}{\textit{Var}}

\newcommand{\CC}{\mathcal{C}}
\newcommand{\RR}{\mathcal{F}}

\newcommand{\OO}{\mathcal{O}}

\newcommand{\srewrites}[1]{\stackrel{#1}{\longmapsto}}

\newcommand{\into}{\ensuremath{\,\rfloor}\,}

\newcommand{\phole}{\square}

\newcommand{\conc}{\;\,}
\newcommand{\emptyseq}{\bullet}

\newcommand{\TSV}{\VV_{\TS}}
\newcommand{\ASV}{\VV_{\AS}}
\newcommand{\VV}{\mathcal{V}}

\newcommand{\mapstoSug}{\Rightarrow}
\newcommand{\smapstoSug}[1]{\stackrel{#1}{\Rightarrow}} 
\newcommand{\mapstoWrapSug}{\rightarrow}
\newcommand{\mapstoDesug}{\mapsto}

\newcommand{\red}{\mapstoDesug}

\allowdisplaybreaks[1]

\newcommand{\st}{simple term}
\newcommand{\sts}{simple terms}
\newcommand{\St}{Simple term}
\newcommand{\Sts}{Simple terms}
\newcommand{\short}{\CalculusShortName}
\newcommand{\ov}[1]{\overline{#1}}
\newcommand{\set}[1]{\{#1\}}

\newif\ifmc
\mcfalse 
\newcommand{\mc}[1]
{\ifmc{\color{magenta}{#1}}\else{#1}\fi}

\newif\ifeg
\newcommand{\eg}[1]
{\ifeg{\color{blue}{#1}}\else{#1}\fi}

\title{Stochastic Calculus of Wrapped Compartments\thanks{This research is funded by the BioBITs Project (\emph{Converging
Technologies} 2007, area: Biotechnology-–ICT), Regione Piemonte.}}
\author{Mario Coppo$^1$, Ferruccio Damiani$^1$, Maurizio Drocco$^1$,  Elena Grassi$^{1,2}$ and Angelo Troina$^1$
\institute{$^1$Dipartimento di Informatica, Universit\`a di Torino}
\institute{$^2$Molecular Biotechnology Center, Dipartimento di
Genetica, Biologia e Biochimica, Universit\`a di Torino}
\email{\{coppo,damiani,troina\}@di.unito.it $\quad$
maurizio.drocco@gmail.com $\quad$ grassi.e@gmail.com} }

\def\titlerunning{Stochastic Calculus of Wrapped Compartments}
\def\authorrunning{M. Coppo, F. Damiani, M. Drocco, E. Grassi \& A. Troina}

\begin{document}

\maketitle

\begin{abstract}
The Calculus of Wrapped Compartments (\short ) is a variant of the Calculus of Looping Sequences (CLS). While keeping the same expressiveness, \short\ strongly simplifies the development of automatic tools for the analysis of biological systems. The main simplification consists in the removal of the sequencing operator, thus lightening the formal treatment of the patterns to be matched in a term (whose complexity in CLS is strongly affected by the variables matching in the sequences).

We define a stochastic semantics for this new calculus. As an application we model the interaction between macrophages and apoptotic \eg{neutrophils} and a mechanism of gene regulation in \emph{E.Coli}.
\end{abstract}

\section{Introduction}

In computer science, several formalisms have been proposed for the
description of the behaviour of biological systems. Automata-based
models~\cite{ABI01,MDNM00} have the advantage of allowing the direct
use of many verification tools such as model checkers. Rewrite
systems~\cite{DL04,P02,BMMT06} usually allow describing
biological systems with a notation that can be easily understood by
biologists. Compositionality allows studying the behaviour of a
system componentwise. Both automata-like models and rewrite systems
present, in general, problems from the point of view of
compositionality, which, instead, is in general ensured by process
calculi, included those commonly used to describe biological
systems~\cite{RS02,PRSS01,Car05}.


The Calculus of Looping Sequences~\cite{BMMT06} (CLS) has been
introduced to describe microbiological systems, such as cellular
pathways, and their evolution. The terms of the calculus are
constructed by basic constituent elements and operators of
sequencing, looping, containment and parallel composition. The
looping operator allows tying up the ends of a sequence, thus
creating a circular sequence which can represent a membrane. The
presence of sequences and loops, however, makes more difficult to
develop a correct and efficient implementation of the associated
simulation tools. The main difficulties come from
the complexity of the pattern matching algorithms which should take
sequences (and all their subsequences) into account. Moreover, the
use of sequences and sequence variables seems prone to subtle semantic
ambiguities in rules definition.

In this paper we propose the \emph{Calculus of
Wrapped Compartments} (\short\ for short), a variant of CLS suitable for
efficient implementation. In \short , starting from an alphabet of
atomic elements, we build multisets of elements and compartments. We
are able to localise elements by compartmentalisation and we allow
to specify the structure of the compartment (e.g., detailing the
elements of interest on a membrane). The evolution of the system is
driven by a set of rewrite rules modelling the reactions of
interest. The main difference with respect to CLS consists in
removing the sequencing operator and using multisets (instead of
ordered sequences) of atoms to represent membranes. Such a choice,
while preserving the Turing completeness of the calculus, strongly
simplifies the pattern matching algorithm \mc{and the underlying
basic theory}.

In our previous experience with biological models we realized that
sequences were not used in many situations: cellular pathways are
usually represented with a greater abstraction than the one which
uses amino acids and nucleotides to depict proteins and genes -
moreover representing the primary sequence of a protein without
considering secondary and tertiary structure is seldom useful. CLS
used looping sequences to represent membranes, but we believe that
multisets of atomic elements are a more natural representation,
which reflects the fluid mosaic model~\cite{SN72}. A similar approach
is taken in the CLS+ model (see~\cite{BMMT07}), a variant of CLS
in which the looping operator can be applied to a parallel composition of
sequences. 

We provide \short\ with a stochastic operational semantics from
which a continuous time Markov chain can be build. With respect
to~\cite{BMMTT08}, in which a constant rate is associated to the
rewrite rules, we allow a more flexible treatment of the dynamics.
In particular, we associate a function to each rewrite rule
according to which the rate of the transition can be computed
depending on the initial and the target terms. We show how the law
of mass action can be encoded in our stochastic semantics thus
following the standard Gillespie's approach~\cite{G77}.

\paragraph{Summary.} In Section~\ref{CWM_formalism} we introduce the new formalism and show its Turing completeness. In Section~\ref{ModGuid} we provide some guidelines for the modelling of common biological events. We give some detail on how the interaction between macrophages and apoptotic \eg{neutrophils}, and a mechanism of gene regulation can be described in the new calculus. In Section~\ref{Stoch-CWM_formalism} we present a stochastic semantics of the calculus. In Section~\ref{pho}, we report some simulations regarding the gene regulation mechanism introduced in Section~\ref{ModGuid}. Finally, in Section~\ref{conc} we draw some conclusions.

\section{The Calculus of Wrapped Compartments}\label{CWM_formalism}

In this section we present the nondeterministic Calculus of Wrapped Compartments (\short ).

\subsection{Terms and Structural Congruence} \label{termCongr}
A \emph{term} of the \short\ calculus is intended to represent a
biological system. A \emph{term} is a sequence of \emph{\st s}.
\Sts, ranged over by $t$, $u$, $v$, $w$, are built by means of the \emph{compartment}
constructor, $(-\into -)$, from a set
$\AT$ of \emph{atomic elements} (\emph{atoms} for short), ranged
over by $a$, $b$, $c$, $d$. The syntax of {\st s} is given at the
top of Figure~\ref{fig:CWM-syntax}. We write $\overline{t}$ to
denote a (possibly empty) sequence of \sts\ $t_1\mydots t_n$.
Similarly, with $\overline{a}$ we denote a (possibly empty)
sequence of atoms. The set of simple terms will be denoted by
$\TT$. The set of terms (sequences of simple terms) and the set of
sequences of atoms will be denoted by $\TS$ and $\AS$,
respectively. Note that $\AS \subseteq \TS$.

A term $\overline{t}=t_1\mydots t_n$ should be understood as the
multiset containing the simple terms $t_1,\ldots,t_n$. Therefore, we
introduce a relation of structural congruence, following a standard
approach in process algebra. The \short\ \emph{structural
congruence} is the least equivalence relation on terms satisfying
the rules given  at the bottom of Figure~\ref{fig:CWM-syntax}. From
now on we will always consider terms modulo structural
congruence.\footnote{In the implementation of the derived tools it
will be useful to consider a normalized representation of these terms
suitable for efficient manipulation. In the description of the
calculus given here however we will ignore implementation issues.}
Then a \st\ is either an atom or a compartment $(\overline{a}\into
\overline{t})$ consisting of a \emph{wrap} (represented by the
multiset of atoms $\overline{a}$) and a \emph{content} (represented
by the term $\overline{t}$).
%
We write the empty multiset as
$\emptyseq$ and denote the union of two multisets $\overline{u}$
and $\overline{v}$ as $\overline{u}\conc\overline{v}$.


\begin{figure}[t]
\hrule $\;$ \\
\begin{tabular}{l}
 \textbf{\St s syntax}
\\
 $
 \begin{array}{lcl}
 \\
  t & \;\qqop{::=}\; & a \!\agr\! (\overline{a}\into\overline{t})
   \\
   \\
 \end{array}
 $
\end{tabular}
\\
\hrule $\;$ \\
\textbf{Structural congruence}
\\
$
\begin{array}{l}
\\
\overline{t} \conc u \conc w \conc \overline{v} \equiv \overline{t}
\conc w \conc u \conc \overline{v} ~~~~~~~~~~~~~~
  \mbox{ if } \; \ov{a}\equiv\ov{b}~~\text{and}~~\ov{t}\equiv\ov{u} ~~\mbox{ then }~~
	(\ov{a}\into\ov{t})\equiv(\ov{b}\into\ov{u})
%
%
\\
\\
\end{array}
$
\hrule
 \caption{ \CalculusShortName\ term syntax and structural congruence rules.}
\label{fig:CWM-syntax}
\end{figure}


%

\begin{figure}[t]
\begin{center}
\begin{minipage}{0.98\textwidth}
\begin{center}
\includegraphics[height=30mm]{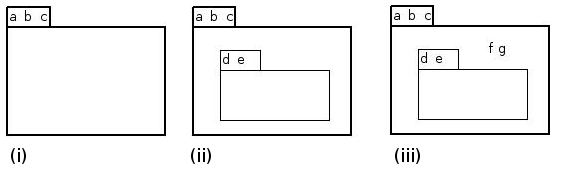}
\end{center}
\vspace{-0.3cm}
\caption{(i) represents $(a \conc b \conc c \into \emptyseq)$; (ii) represents $(a
\conc b \conc c  \into (d \conc e \into \emptyseq))$; (iii) represents $(a
\conc b
\conc c \into (d \conc e \into \emptyseq) \conc f \conc g)$.}
\label{fig:example CWM}
\end{minipage}
\end{center}
\end{figure}

An example of term is $a \conc b \conc (c \conc d \into e \conc
f)$ representing a multiset consisting of two atoms $a$ and $b$
(for instance two molecules) and a compartment $(c \conc d \into e
\conc f)$ which, in turn, consists of a wrap (a membrane) with two
atoms $c$ and $d$ (for instance, two proteins) on its surface, and
containing the atoms $e$ (for instance, a molecule)
 and $f$ (for instance a DNA strand). See Figure~\ref{fig:example CWM} for some
graphical representations.

\subsection{Systems and Reduction Semantics}

Rewrite rules are defined as pairs of terms, in which
the left term characterizes the portion of the system in which the
event modelled by the rule can occur, and the right one describes
how that portion of the system is changed by the event.

In order to formally define the rewriting semantics, we introduce
the notion of open term (a term containing variables) and pattern
(an open term that may be used as left part of a rewrite rule). In
order to respect the syntax of terms, we distinguish between ``wrap
variables'' which may occur only in compartment wraps (and can
be replaced only by multisets of atoms) and ``term variables'' which
may only occur in compartment contents or at top level (and can be replaced
by arbitrary terms).
%
%
Therefore, we assume a set of \emph{term variables}, $\TSV$, ranged over
by $X,Y,Z$, and a set of \emph{wrap variables}, $\ASV$, ranged
over by $x,y,z$. These two sets are disjoint. We denote by $\VV$
the set of all variables $\TSV \cup \ASV$, and with $\rho$ any
variable in $\VV$.
%

\begin{definition}[Open Terms and Patterns]\label{DefCWM_OT and Pat} $\;$\\
\vspace{-0.3cm}
\begin{itemize}
\item[(i)] \emph{Open terms} are terms
	which may contain occurrences of wrap variables in compartment
	wraps and term variables in compartment contents or at top level.
	They can be seen as multisets of \emph{simple open terms}. More formally, open terms, ranged over by $\RightPat$ and simple open terms, ranged over by $o$, are defined in the following way:
$$
\begin{array}{lcl}
\RightPat   & \qqop{::=} & \overline{o} \\
o           & \qqop{::=} & a \agr X \agr   (\overline{a}\conc \overline{x}
  \into \overline{o} )
\end{array}
$$
\item[(ii)]  An open term is \emph{linear} if each variable occurs at most
	once in it.
 \item[(iii)]\emph{Simple patterns}, ranged over by $\Pat$, are the linear
	 open terms defined in the following way:
\begin{align*}
\Pat\; & \;\qqop{::=}\; t
 \agr
 (\overline{a} \conc x \into \overline{\Pat}\conc X)
\end{align*}
\normalsize where $t$ is an element of $\TT$, $\overline{a}$ is an
element of $\AS$, $x$ is a variable in $\ASV$, $\overline{\Pat}$ is a possibly
empty multiset of simple patterns and $X$ is a variable in
$\TSV$. We denote with $\PP$ the set of
	 simple patterns.
\item[(iv)] \emph{Patterns}, ranged over by $P$, are the linear
	 open terms defined in the following way:
\begin{align*}
 P & \;\qqop{::=}\; \Pat \conc \overline{\Pat} \conc X
 \\
\end{align*}
\normalsize where $\Pat~\overline{\Pat}$ is a nonempty multiset of
simple patterns and $X$ is an element of $\TSV$. We denote with $\LeftPP$ the
set of patterns.
 \end{itemize}
\end{definition}

   %

An \emph{instantiation} is a partial function $\sigma : \VV \rightarrow
\TS$. An instantiation must preserve the type of variables, thus for $X \in
\TSV$ and $x \in \ASV$ we have $\sigma(X) \in \TS$ and $\sigma(x) \in \AS$,
respectively.  Given $\RightPat \in \OT$, with $\RightPat \sigma$ we denote
the term obtained by replacing each occurrence of each variable $\rho \in
\VV$ appearing in $\RightPat$ with the corresponding term $\sigma(\rho)$.
With $\Sigma$ we denote the set of all the possible instantiations, and,
given $\RightPat \in \OT$, with $\VarOf(\RightPat)$ we denote the set of
variables appearing in $\RightPat$. Now we can define rewrite rules.

\begin{definition}[Rewrite Rules]\label{Def Rewrite Rules}
A \emph{rewrite rule} is a pair of a pattern  $\LeftPat$ and an
open term $\RightPat$, denoted with $\LeftPat \! \mapstoDesug \!
\RightPat$, such that $\VarOf(\RightPat) \subseteq
\VarOf(\LeftPat)$.
\end{definition}

A rewrite rule $\LeftPat \! \mapstoDesug \! \RightPat$ states that
a term $\LeftPat \sigma$, obtained by instantiating variables in
$\LeftPat$ by some instantiation function $\sigma$, can be
transformed into the term $\RightPat\sigma$. A set of rewrite
rules define a notion of reduction between terms of \short\ which
are intended to represent the (nondeterministic) behaviour of the
represented system.

\begin{definition}
 A \emph{\short\ system} over a set  $\AT$
of atoms is represented by a set $\RR_\AT$ ($\RR$ for short when
$\AT$ is understood) of rewrite rules over $\AT$.
\end{definition}

We define the semantics of \CalculusShortName\ as a transition system in
which states correspond to terms and transitions correspond to rule
applications. To this aim we need to define the notion of contexts (i.e. terms containing a hole).

\begin{definition}[Contexts]\label{Def Contexts} \emph{Contexts} $C$ are defined as:
$$
 C  \;\qqop{::=}\; \phole \agr  (\overline{a} \into C) \conc \overline{t}
$$
where $\overline{a} \in \AS$ and $\overline{t}\in \TS$. The context
$\phole$ is called the \emph{empty context}. We denote with $\CC$ the set
of contexts.
\end{definition}

By definition, every context contains a single hole $\phole$. Let us assume
$C,C'\in \CC$. With $C[\overline{t}]$ we denote the term
obtained by replacing $\phole$ with $\overline{t}$ in $C$; with
$C[C']$ we denote context composition, whose result is the context
obtained by replacing $\phole$ with $C'$ in $C$. For example, given
$C=(a \conc b \into \phole)\conc  i \conc l$, $C'= (c\conc  d \into \phole)\conc  g \conc h$ and $\overline{t}= e \conc f$,
we get $C[C'[\overline{t}]]= (a \conc b \into (c \conc d \into e \conc f) \conc g \conc h )\conc  i \conc l$.
Note that context
holes take the place either of the whole term or of the whole
content of a compartment. This allows to make
context unambiguous in the following sense. Let's extend the notion
of subset (denoted as usual as $\subseteq$) between terms
interpreted as multisets.

\begin{proposition}[Uniqueness]\label{unicity}
For any term $\ov{t}$ if the term $\ov{u}$  occurs in $\ov{t}$
within a compartment content or at top level, then there are, modulo $\equiv$, a unique
context $C$ and a unique term $\ov{t'}$  such that $\ov{t} = C[\ov{t'}]$ and
$\ov{u}\subseteq \ov{t'}$.
\end{proposition}

Given a \short\ system $\RR$ the associated notion of
reduction is then defined in the following way.
\begin{definition}[Reduction]\label{DefCWM Reduction}
Given a \short\ system $\RR$, the
	\emph{$\RR$-reduction relation} between terms of
\CalculusShortName\ is
	defined by the following rule:
$$
\frac{ \LeftPat \mapstoDesug \RightPat \in \RR  \qquad \sigma \in
\Sigma  \qquad \overline{t} = \LeftPat\sigma \qquad \overline{u} = \RightPat\sigma \qquad C\in \CC}
	{C[\,\overline{t}\,] \red_\RR C[\overline{u}]}
$$
 closed under structural congruence, i.e.:
$$ \overline{u} \equiv \overline{u'}~~\text{and}~~\overline{u'} \red_\RR
\overline{v'}~~\text{and}~~\overline{v'} \equiv \overline{v}
~~~~~\Rightarrow ~~~~~
	\overline{u}\red_\RR \overline{v}
$$
  We will write simply $\red$ instead of $\red_\RR$ when $\RR$ is
  understood.
\end{definition}

%
\begin{remark} \label{DifferentOutcomes}   $\;$\\
\vspace{-0.3cm}
 \begin{itemize}
\item[(i)]
Taken a pattern $P = \Pat \conc \overline{\Pat} \conc
X$ (representing the reactants of the reaction that will be simulated) the crucial
point for determining an application of $P$ to the whole biological
ambient $\ov{t}$ is the choice of the occurrences of simple terms
matching with $\Pat \conc \overline{\Pat}$ (determining the
compartment in which this reaction will take place). By Proposition
\ref{unicity} and the linearity condition both
the context $C$ (such that $\ov{t} \equiv C[\ov{t'}]$)  and the term
replacing $X$ (the part of $\ov{t'}$ that does not match $\Pat \conc
\overline{\Pat}$) are uniquely determined. This will play a role in the implementation of the stochastic calculus.
\item[(ii)]  There can be, in general, many different substitution such that
$\LeftPat\sigma \equiv \overline{t}$ but not all of them produce
the same  term $\RightPat\sigma$ (modulo $\equiv$). This is not particularly relevant for the case of
qualitative systems but we will need to take it into account in the
stochastic formulation.
\end{itemize}
\end{remark}

\begin{notation}$\;$\\
\vspace{-0.3cm}
\begin{itemize}
\item[(i)] We write
$\Pat\conc\overline{\Pat}\mapstoSug\RightPat$ as short for
$\Pat\conc\overline{\Pat}\conc X \mapstoDesug \RightPat\conc X$,
where the variable $X$ does not occur in
$\VarOf(\Pat\conc\overline{\Pat})\cup\VarOf(\RightPat)$.

\item[(ii)] We write $a\conc\overline{a}\mapstoWrapSug\overline{b}$ as short for the
rewrite rule $(a\conc\overline{a}\conc x \into Y) \conc Z
 \mapstoDesug (\overline{b}\conc x \into Y) \conc Z$,
 where the variables $x,Y,Z$ are distinct.
 \end{itemize}
\end{notation}

The  \CalculusShortName\ is Turing Complete. In the following we sketch how
Turing Machines can be simulated by  \CalculusShortName\ models.

\begin{theorem}[Turing Completeness] The class of  \CalculusShortName\ models is Turing complete.
\end{theorem}
\noindent \emph{Proof:} (Sketch).
A Turing machine $T$ over an alphabet $\Sigma\cup\set{\blank}$
(where $\blank$ represent the blank) with a set $Q$ of states can be
simulated by a system $\RR_T$ of \CalculusShortName\ in the following way. \\
 Take $\AS = Q \cup\Sigma \cup \set{\blank, l, r}$, where $l$ and $r$ are
special symbols to represent the left and right ends of the tape.
The tape of the Turing machine can be represented by a sequence of
nested compartments whose wraps consist of a single atom (representing a symbol of the tape).
The content of each compartment defined in this way represents a
right suffix of the written portion of the tape, while the atom on the wrap represents
its initial (w.r.t. the suffix) symbol. In each term representing a tape there is
exactly one compartment which contains a state (the present state).
 For example, the term $(l \into ( a \into q ~(b \into
(r\into \emptyseq))))$ represents the configuration in which the
tape is $...\blank,a,b,\blank....$ (the remaining positions are
blank), the machine is in state $q$ and the head is positioned on
$b$. Rewrite rules are then used to model the machine evolution. We
can define rules creating new blanks when needed (at the tape ends)
to mimic a possibly infinite tape. For instance a transition $(q, b)
\rightarrow (q',c,right)$ (meaning that in state $q$ with a $b$ on
the head the machine goes in state $q'$ writing $c$ on the tape and
moving the head right) is represented by the rules\footnote{In these rules, in a Turing Machine simulation, the variables $X$ and $Y$ are always instantiated with the empty multiset }:
	$$\begin{array}{ll}
   (1) & q~(b~y\into (x \into Z)~ Y )~ X \red  (c~y \into q' ~ (x \into Z) ~ Y) ~ X \\
	(2)&   q~(b~y\into (r\into Z)~Y) ~ X \red  (c~y \into
	q'~(\blank\into (r \into Z ))~Y)~X
	\end{array}$$
The second rule represents the case that b is the rightmost non blank
symbol and so a new blank must be introduced in the simulation.
\footnote{Note that rule (1) could be applied also in this case but
it would lead the system in a configuration (containing a subterm
$q'~ (r\into t)$ for some t) from which no further move could be
possible. However situations of this kind can be easily avoided with
a little complication of the encoding.} By construction, the system $\RR_T$ correctly represents the
behaviour of $T$.\qed

\section{Modelling Guidelines}\label{ModGuid}

\begin{table}[t]
\begin{center}
\begin{tabular}{|l|l|}
\hline
{\bf Biomolecular Entity} & {\bf CWM term} \\
\hline
\hline
Elementary object (genes, domains, etc...) & Atoms \\
\hline
Molecular population
    & Term multiset \\
\hline
Membrane
    & Atom multiset \\
\hline
\end{tabular}
\end{center}
\caption{Guidelines for the abstraction of biomolecular entities
into \short .}\label{tab:guidelines-entities}
\end{table}

\begin{table}[t]
\begin{center}
\begin{tabular}{|l|l|}
\hline
{\bf Biomolecular Event} & {\bf Examples of \CalculusShortName{ }Rewrite Rules} \\
\hline
\hline
State change &
    $a \mapstoSug b$ \\
\hline
Complexation &
    $a \conc b \mapstoSug c$ \\
\hline
Decomplexation &
    $c \mapstoSug a \conc b$ \\
\hline
Catalysis &
    $c\conc\Pat\conc\overline{\Pat}\mapstoSug c\conc\RightPat$
    where $\Pat\conc\overline{\Pat}\mapstoSug\RightPat$ is the catalyzed event \\
\hline
State change on membrane &
    $ a \mapstoWrapSug b $ \\
\hline
Complexation &
    $ a \conc b \mapstoWrapSug c $ \\
on membrane
    &  $ a \conc (b \conc x \into X) \mapstoSug (c \conc x \into X) $ \\
    &  $ (b \conc x \into a \conc X) \mapstoSug (c \conc x \into X) $ \\
\hline
Decomplexation &
    $ c \mapstoWrapSug a \conc b $ \\
on membrane
    &  $ (c \conc x \into X) \mapstoSug  a \conc (b \conc x \into X) $ \\
    &  $ (c \conc x \into X) \mapstoSug  (b \conc x \into a \conc X) $ \\
\hline
Catalysis on membrane &
    $c\conc a\conc\overline{a}\mapstoWrapSug c \conc \overline{b}$
    where $a\conc\overline{a}\mapstoWrapSug\overline{b}$ is the catalyzed event \\
\hline
Membrane crossing &
    $ a \conc (x \into X) \mapstoSug  (x \into a \conc X) $\\
    & $ (x \into a \conc X) \mapstoSug  a \conc (x \into X) $\\
\hline
Catalyzed&  $ a \conc (b \conc x \into X) \mapstoSug  (b \conc x \into a \conc X) $\\
membrane crossing & $ (b \conc x \into a \conc X) \mapstoSug  a \conc (b \conc x \into X) $\\
\hline
Membrane joining &
    $ a \conc (x \into  X) \mapstoSug  (a \conc x \into X) $\\
    & $ (x \into a \conc X) \mapstoSug (a \conc x \into X)$\\
\hline
Catalyzed &
    $ a \conc (b\conc x \into X) \mapstoSug (a \conc b \conc x \into X) $ \\
membrane joining &
    $ (b\conc x \into a \conc X) \mapstoSug (a \conc b \conc x \into X) $ \\
    & $ (x \into a \conc b \conc X) \mapstoSug (a \conc x \into b \conc X) $\\
\hline
Membrane fusion &
    $ (a \conc x \into X) \conc (b \conc y \into Y) \mapstoSug (a \conc x \conc b \conc y \into X \conc Y) $ \\
\hline
Vesicle dynamics &
    $ (x \into X \conc (a \conc y \into Y)) \mapstoSug Y \conc (a \conc y \conc x \into X) $ \\
\hline
\end{tabular}
\normalsize
\end{center}
\caption{Guidelines for the abstraction of biomolecular events into
\short .}\label{tab:guidelines-events}
\end{table}

In this section we will give some explanations and general hints
about how \CalculusShortName{ }could be used to represent the
behaviour of various biological systems.

In rewrite systems, such as \CalculusShortName, entities
are usually represented by terms of the rewrite
system, and events by rewrite rules. We have already introduced the
biological interpretation of our terms in Section~\ref{termCongr}.
Here we will go deeper and describe the most straightforward
abstraction that can be used to define terms and rules.

First of all, we should select the biomolecular entities of
interest. Since we want to describe cells, we consider molecular
populations and membranes. Molecular populations are groups of
molecules that are in the same compartment of the cells and inside them.
As we have said before, molecules can be of many types: we classify them as
proteins, chemical moieties and other molecules.

Membranes are considered as elementary objects: we
do not describe them at the level of the phospholipids they are made of.
The only interesting properties of a membrane are that it may have a
content (hence, create a compartment) and that in its phospholipid
bilayer various proteins are embedded, which act for example as transporters
and receptors. Since membranes are represented as multisets of the embedded
structures, we are modelling a fluid mosaic in which the membranes become
similar to a two-dimensional liquid where molecules
can diffuse more or less freely~\cite{SN72}.

Now, we select the biomolecular events of interest. The simplest
kind of event is the change of state of an elementary object. Then,
there are interactions between molecules: in particular
complexation, decomplexation and catalysis.
Interactions could take place between simple molecules, depicted
as single symbols, or between membranes and molecules: for
example a molecule may cross or join
a membrane. Finally, there are also interactions between membranes: in
this case there may be many kinds of interactions (fusion, vesicle
dynamics, etc\ldots).

The guidelines for the abstraction of biomolecular entities and
events into \CalculusShortName{ }are given in Table~\ref{tab:guidelines-entities} and
Table~\ref{tab:guidelines-events}, respectively.
Biomolecular events are associated
with rewrite rules. In the table we give some examples of
rewrite rules for different kinds of events. The list has not the
ambition of being complete but to give
some general hints on rules definition.

\subsection{Modelling Examples}\label{Sec:examples}
We will now present two examples of possible \CalculusShortName{ }simulations,
in order to testify that our calculus is
able to express a broad range of biological situations even if the possibility to represent
sequences, which is a computational burden, is removed from its syntax.

\paragraph{Macrophages.}

All the biological problems involving cell interactions are feasible.
One could for example represent macrophages and their interaction with the apoptotic
cells: in order to ``clean up'' the remains of death cells they have to recognize them amongst
the alive ones and then remove them by a process called phagocytosis, which consists
in the internalization of the death cell in a membrane compartment, called phagosome.
The phagosome inside the macrophage will then join another membrane compartment
called the lysosome in order to start the degradation of the ingested cellular components.

\begin{figure}[t]
\begin{center}
\begin{minipage}{0.98\textwidth}
\begin{center}
\includegraphics[width=130mm]{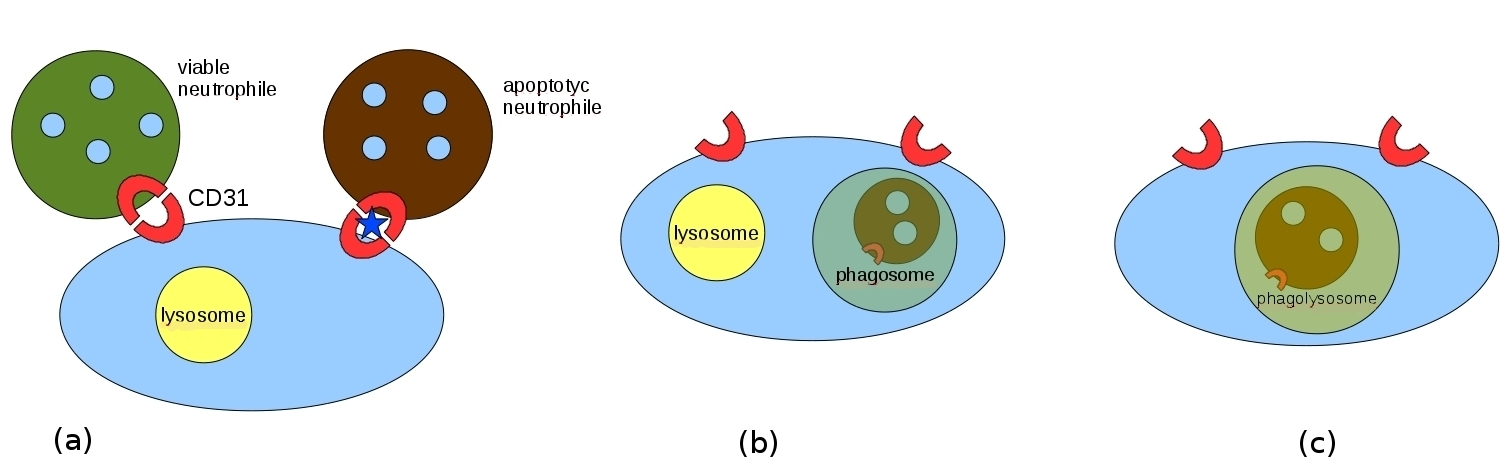}
\end{center}
\vspace{-0.5cm}
\caption{Neutrophils phagocytosis: a) a macrophage distinguishes between apoptotic and living neutrophils, b) the death neutrophile is ``eaten'' and embedded in a phagosome, c) the phagosome fuses with a lysosome.}
\label{fig:phago}
\end{minipage}
\end{center}
\end{figure}

To give the rules that could represent this process in \CalculusShortName{ }we
decided to concentrate on the removal of neutrophils that have undergone
apoptosis: removing them before their membrane collapses is crucial because they
contain some lytic enzymes that, if released in the organism, would lead to tissue damages~\cite{FK03,CGX95}.
The mechanism which mediates the recognition of apoptotic neutrophils is not
perfectly clear but we decided to reproduce the one described in~\cite{VWATR07,BHR02} which
suggest that neutrophils and macrophages interact via CD31 and that somehow
the CD31 of vial neutrophils is capable of ``escaping'' the phagocytosis.

We will represent a macrophage with a lysosome and a CD31 exposed on its membrane in this way (where M
is a marker to distinguish macrophages from other cells and $lyso$ represents a lysosome containing some lytic enzymes):\footnote{Our calculus allows to represent with simple atoms some higher level information such as that represented by marker names.}
\begin{equation}
(\mbox{CD31} \conc \mbox{M}\into (lyso \into lyticEnz) \conc innerM ) \nonumber
\end{equation}
A living neutrophile (where N is a marker for neutrophils and V indicates that this is a viable neutrophile) is represented by:
\begin{equation}
(\mbox{CD31} \conc \mbox{V} \conc \mbox{N} \into innerL) \nonumber
\end{equation}
An apoptotic one (where A indicates its apoptotic state) is represented by:
\begin{equation}
(\mbox{CD31} \conc \mbox{A} \conc \mbox{N} \into innerA) \nonumber
\end{equation}
where $innerM$, $innerL$ and $innerA$ denote the internal material of a macrophage, a living neutrophile and an apoptotic one, respectively.

Now a fairly simple set of rules is sufficient to represent the whole process.

The binding of a neutrophile and a macrophage mediated by their CD31 molecules is represented by encapsulating the interacting cells within a new compartment wrapped with the marker I:
\begin{equation}
(\mbox{CD31} \conc \mbox{M} \conc x \into  X) \conc (\mbox{CD31} \conc \mbox{N} \conc y \into Y) \mapstoSug ( \mbox{I} \into (\mbox{CD31} \conc \mbox{M} \conc x \into  X) \conc (\mbox{CD31} \conc \mbox{N} \conc y \into Y) ) \nonumber
\end{equation}
After the contact the viable neutrophils are released:
\begin{equation}
( \mbox{I} \conc z \into (\mbox{CD31} \conc \mbox{M} \conc x \into  X ) \conc (\mbox{CD31} \conc \mbox{N} \conc \mbox{V} \conc y \into Y) \conc Z ) \mapstoSug  (\mbox{CD31} \conc \mbox{M} \conc x\into  X )\conc (\mbox{CD31} \conc \mbox{N} \conc \mbox{V} \conc y \into Y) \nonumber
\end{equation}
While the apoptotic ones are phagocyted ($phago$ represents the phagosome which contains the phagocyted neutrophile):
\begin{equation}
( \mbox{I} \conc z \into (\mbox{CD31} \conc \mbox{M} \conc x \into  X) \conc (\mbox{CD31} \conc \mbox{N} \conc \mbox{A} \conc y \into Y) \conc Z ) \mapstoSug (\mbox{CD31} \conc \mbox{M} \conc x \into  (phago \into (\mbox{CD31} \conc \mbox{N} \conc \mbox{A} \conc y \into Y)) \conc X) \nonumber
\end{equation}
The phagosome can join the lysosome to start hydrolysis ($phagolyso$ represents the phagolysosome derived from the fusion):
\begin{equation}
 (lyso \conc x \into lyticEnz \conc X) \conc (phago \conc z \into (\mbox{CD31} \conc \mbox{N} \conc \mbox{A} \conc y \into Y) \conc Z)  \mapstoSug
(phagolyso \conc x \conc z \into (\mbox{CD31} \conc \mbox{N} \conc \mbox{A} \conc y \into Y) \conc lyticEnz \conc X \conc Z)  \nonumber
\end{equation}

This should show that interactions involving cells of the immune system can be represented in \short\ with little effort at different levels
of detail: modelling some aspects of the complex network that drives
immunological defence could be useful to study many human pathologies and involved processes, like in this example
about inflammation and how it stops.
\begin{figure}[t]
\begin{center}
\begin{minipage}{0.98\textwidth}
\begin{center}
\includegraphics[width=130mm]{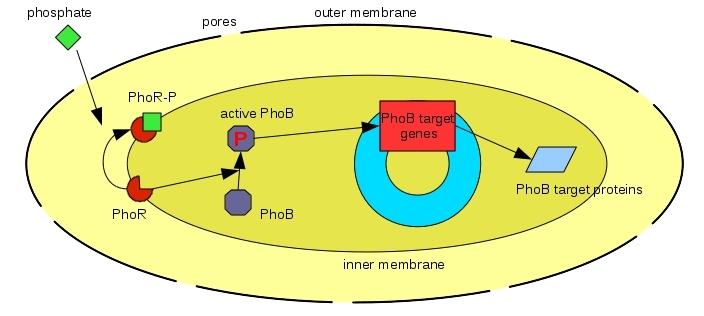}
\end{center}
\vspace{-0.5cm}
\caption{The Pho regulation system.}
\label{fig:pho}
\end{minipage}
\end{center}
\end{figure}

\paragraph{Phosphate regulation.}
Switching to a completely different field we would like to describe
a mechanism of gene regulation fairly well defined in E.Coli and
other bacteria, which is based on a two component system: \emph{(i)} the
sensor, which can react to changes in the environment and interact
with the other component; \emph{(ii)} the response regulator, which is capable
of determining, when activated by the sensor, the reactions
necessary to cope with the new environmental conditions. This schema
is common to drive gene regulation in bacteria: the activated
response regulator will act as a transcriptional activator.

We will model the system related to environmental phosphate concentration, where the sensor is
a transmembrane protein located on the inner membrane of the bacteria, called PhoR, and the response regulator
is a cytoplasmatic protein which can be activated by phosphorylation and is called PhoB.
Environmental phosphate will freely diffuse in the periplasmic space (between the two
bacterial membranes) thanks to the outer membrane
pores and there it will bind to PhoR. When bound by phosphate PhoR will be silent, while it shows
a kinase activity on PhoB when it is unbound; therefore low levels of phosphate will leave many PhoR
unbound and able to activate PhoB, while high levels will lead to a smaller fraction of transmembrane
unbound PhoR and thus to little activation of PhoB. When phosphorylated, PhoB activates the transcription of
many genes aimed at helping the bacteria survive in low phosphate conditions~\cite{MCB00}.
These mechanisms have been recently described as related to the pathogenic potential~\cite{FS08} of some bacteria
and therefore studying them could also have some importance in medicine.

We will now present the set of rules needed to model this regulation system; in the following
we will refer to the outer membrane pores as $pore$, to phosphate as $Pi$,
to the phosphate bound and unbound PhoR as  $PhoR$ and $PhoRP$,
to the inactive and active PhoB as $PhoB$ and $PhoBP$, to its target genes
as $PhoGenes$ and to the proteins that they encode as $PhoProt$.

We may represent an E.Coli with the following \short\ term:
\begin{equation}\nonumber
(pore  \into (\ov{PhoR} \conc \ov{PhoRP} \into \ov{PhoB} \conc \ov{PhoBP} \conc PhoGenes ) )
\end{equation}
where, here, a sequence $\ov{PhoR}$ represents a multiset of $PhoR$ atoms (similar for the other atoms).
Two rules describe the phosphate diffusion in the periplasmic space:
\begin{equation}\nonumber
\begin{aligned}
& Pi \conc (pore \conc x \into X) \mapstoSug (pore \conc x \into Pi \conc X) \\
& (pore \conc x \into Pi \conc X) \mapstoSug Pi \conc (pore \conc x \into X)
\end{aligned}
\end{equation}
Two rules for the PhoR phosphate binding:
\begin{equation}\nonumber
\begin{aligned}
& Pi \conc (PhoR \conc x \into X) \mapstoSug (PhoRP \conc x \into  X) \\
& (PhoRP \conc x \into  X) \mapstoSug Pi \conc (PhoR \conc x \into X)
\end{aligned}
\end{equation}
PhoB phosphorylation:
\begin{equation}
(PhoR \conc x \into  PhoB \conc X) \mapstoSug (PhoR \conc x \into PhoBP \conc X) \nonumber
\end{equation}
High level rules that represent the translation and traduction to proteins of the PhoB target genes and these
proteins degradation:
\begin{equation}\nonumber
\begin{aligned}
& PhoBP \conc PhoGenes \mapstoSug PhoBP \conc PhoGenes \conc PhoProt \\
& PhoProt \mapstoSug \emptyseq
\end{aligned}
\end{equation}
One last rule represents the bound form of PhoR phosphatase activity on PhoBP~\cite{CHM03}, which
is a negative regulating loop of the whole mechanism:
\begin{equation}
(PhoRP \conc x \into  PhoBP \conc X) \mapstoSug (PhoRP \conc x \into PhoB \conc X) \nonumber
\end{equation}

In Section~\ref{pho} we will give some details about simulations of this model.

\section{\mc{The \SCalculusLongName}}\label{Stoch-CWM_formalism}

In order to make the formal framework suitable to model quantitative
aspects of biological systems we must associate to each rule a numerical
parameter (the rule \emph{rate}) which determines (in a stochastic sense)
the time spent between two successive interactions and the frequency with
which each interaction will take place in order to represent faithfully the
system evolution. The operational semantics will be defined by
incorporating the rule rates in a stochastic framework along the lines of
the one presented by Gillespie in \cite{G77}. The same approach has been
incorporated, for example, in other simulators as the ones for the
stochastic $\pi$-calculus \cite{P95,PRSS01}.

In this framework the rule rate is used as the parameter of an exponential
distribution modelling the time spent between two occurrences of the
considered chemical reaction. The use of exponential distributions to
represent the (stochastic) time spent between two occurrences of chemical
reactions allows describing the system as a Continuous Time Markov Chain
(CTMC), and consequently allows verifying properties of the described
system analytically and by means of stochastic model checkers.

The basic choice for determining the rule rate of a reaction (suggested
also by Gillespie's algorithm) is to obtain it by multiplying the kinetic
constant of the reaction by the number of possible combinations of
reactants that may occur in the system, representing in this way the law of
mass action.

In our calculus we will represent a reaction rate as a
function of the context in which the reaction takes place. This allows to
tailor the reaction rates on the specific characteristics of the system, as for
instance when representing nonlinear reactions as it happens for Michaelis--Menten kinetics. The standard mass action
rate can be describes as a particular choice of the rate
function (see below). A similar approach is used in~\cite{DGT09b} to model reactions with inhibitors and catalysers in a single rule.

Obviously some care must be taken in the choice of the rate function: for
instance it must be complete (defined on the domain of the
application) and nonnegative. This properties are enjoyed by the function
representing the law of mass action.

\begin{definition}
  A \emph{stochastic} rewrite rule is a triple $(P, R, f)$, denoted
$P\srewrites{f} R$, where $(P,R)$ is a rewrite rule and $f:\TS \times \TS
\rightarrow \mathbb{R}^{\geq 0}$ is the \emph{rate function} associated to
the rule.\footnote{The value $0$ in the codomain of $f$ models the situations in which
the given rule cannot be applied, for example when the particular
environment conditions forbid the application of the rule.}
\end{definition}

We can then define, as in Definition \ref{DefCWM Reduction}, the
``stochastic'' reduction relation for \short. In this case, we must also
take into account that, by Remark \ref{DifferentOutcomes}, different
instantiations that allow the l.h.s. $P$ of a rule to match a term $\ov{t}$
can produce different outcomes which could determine different rates. This
is shown in Example~\ref{ex:multiterm}. So, in the definition of the rate
function, we must also take into account the term produced by the reaction
after the application of the rule.

\begin{example}\label{ex:multiterm}
Consider the rewrite rule $a\conc (b \conc x \into X) \srewrites{f} (a\conc
b \conc x \into X) $ modelling a catalyzed membrane joining. In this case,
the initial term $ a \conc (b \conc b \into c) \conc (b \into c)$ results
in two different terms depending on which membrane will be joined by the
element $a$; namely $ (a \conc b \conc b \into c) \conc (b \into c)$ and $
(b \conc b \into c) \conc (a \conc b \into c)$. The application of the
rule, moreover, should take into account that the first case is catalyzed by
the presence of two elements $b$ on the membrane, while the second one has
only one catalyzer.
\end{example}

A \emph{Stochastic \short\ systems} $\mathcal{R}_\AT$ ($\mathcal{R}$ for
short) can be defined as in Definition~\ref{DefCWM Reduction} from a set
$\mathcal{R}$ of
stochastic reduction rules over $\AT$. \\


\begin{definition}[Stochastic Reduction]\label{DefCWM SReduction}
 Given a stochastic \short\ system
	$\mathcal{R}$, the induced \emph{stochastic reduction relation} between
terms of
	\CalculusShortName\ is \mc{defined by the following rule}:
$$
\frac{ P \srewrites{f}  O \in \mathcal{R} \qquad\sigma \in \Sigma
		\qquad \ov{t} = P\sigma\qquad \ov{u}= O\sigma
		 \qquad C\in \CC}
	{ C[\ov{t}] \srewrites{f(\ov{t}, \ov{u})}_\mathcal{R}  C[\ov{u}]}
$$
 We will write simply $\srewrites{f}$ instead of $\srewrites{f}_\mathcal{R}$ when $\mathcal{R}$ is
  understood.
\end{definition}

\paragraph{Chemical--like Reactions.}

To define the ``canonical'' function that represents the law of mass
action the application rate of a reduction rule should depend on the
kinetic constant of the reaction and the number of possible
reactants in the considered compartment in which the reaction will
take place. Following an approach similar to that of
\cite{BMMTT08,KMT08}, this can be obtained counting the number of
different ways in which the rule representing a reaction can be
applied in the considered context producing the same outcome. \mc{In
our setting, however, we can formalize the canonical function in a
simpler way.}

To this aim we need a few definitions. Assume to have an infinite set of
\emph{labels} (e.g. the natural numbers) that can be attached to the atomic
elements in $\AT$  to create, for each $a \in \AT$ an infinite set of
distinguished versions $a_1, a_2, a_3,...$ of the element. If $a_i$ is a
labelled atom, we write $a = |a_i|$ for the \emph{support} of
$a_i$.

\begin{definition}\label{labeled}$\;$\\
\vspace{-0.3cm}
\begin{itemize}
\item[(i)] A \emph{labelled term} of \short\ is a term in which some
	labelled atomic element can occur.
\item[(ii)] A \emph{complete} labelled term is a term in which all atomic
	elements are labelled and no two labelled atoms having the same
	support have the same label.
\item[(iii)] If $\ov{t}$ is a labelled term, its \emph{erasure} $|\ov{t}|$ is
	the term obtained from $ \ov{t}$ by erasing all the labels.
\end{itemize}
\end{definition}

\mc{Labels can be applied also to atoms occurring in open terms, but
no label can be associated to a variable.

If $\ov{t}$ is a labelled (possibly open) term a \emph{labeling}
$\ov{t^+}$ of $\ov{t}$ is obtained by assigning labels to the non
labelled atoms of $\ov{t}$ such that $\ov{t^+}$ is a completely
labelled term.\footnote{Note that there are many possible labellings of
$\ov{t}$.} For example, taking $\ov{t} = a_1~a~a_3~b$ a possible
labeling of $\ov{t}$ is  $\ov{t^+} = a_1 ~ a_2 ~ a_3 ~ b_1 $.}

Given a reduction rule $P\srewrites{f}
O$ and terms $\ov{t}, \ov{u}$ such that for some instantiation
$\sigma$ it holds $P\sigma \equiv \ov{t}$ and $O\sigma \equiv \ov{u}$, we
define the canonical rate function $f(\ov{t},\ov{u})$ by the
following steps:
  \begin{enumerate}
  \item take any completely labeled version $\ov{t^+}$ of $\ov{t}$;
  \item let $n$ be the number of distinct instantiations of variables
	  with completely labeled terms such that $P'\sigma\equiv
	  \ov{t^+}$, for some labeling $P'$ of $P$, and $|R\sigma|\equiv
	  \ov{u}$;\footnote{Note that the labeling of the atoms in $P$ is
	  taken only in order to allow matching. It is the choice of the
	  variables the relevant factor.}
  \item define the function $f$ as $f(\ov{t}, \ov{u}) = k\cdot n$ where
	  $k$ is the kinetic constant of the associated reaction.
  \end{enumerate}

We denote  with $f^{can}_k$ the function obtained in this way to
compute the evolution rate following the law of mass action with
kinetic constant $k$.

Note that since the labels of the atomic elements are all different, by
Proposition \ref{unicity} and Remark \ref{DifferentOutcomes} (i)  each
instantiation $\sigma$ such that $P'\sigma\equiv \ov{t^+}$ corresponds to a
distinct combination of the elements that determine the reaction described
by the rule $P\srewrites{} O$. So the number $n$ of substitutions defined
in point 2. corresponds exactly to the number of possible distinct reactants
for the considered reaction.

This number can often be calculated in a rather simple way. For instance in
a rule  $X\;a\;b~\srewrites{}~X\;c$ each instantiation of $X$ that matches any
term containing at least one $a$ and one $b$ corresponds exactly to a pair
of occurrences of $a$ and $b$ (the ones not included in the variable's instantiation). So the total number of
distinct instantiations of $X$ (i.e. of possible reactants) is given
by the product of the number of occurrences of $a$ times the number of
occurrences of $b$.

%

\begin{example}
 Consider the completely labeled term $\ov{t^+}=a_1 ~ a_2 ~ a_3 ~ b_1 $ defined
  above and the qualitative reduction rule $a~a~X\red a ~c ~X$. The l.h.s.
 of the rule can be matched by the instantiations $X = a_1~b_1$,
 $X = a_2, b_1$ and $X = a_3 ~ b_1$. In the first case, the match is
 obtained by taking $P' = a_2 ~ a_3 ~ X$ and so on. So, we have that
 $f^{can}_k(a~a~a~b,~a~a~c~b) = 3\cdot k$ where $k$ is the kinetic
 constant of the reaction.
 \end{example}

 \begin{example}
 Consider again the term $ \ov{t}=a \conc (b \conc b \into c) \conc (b \into c)$ and the rewrite rule  $a\conc (b \conc x \into X) \srewrites{f} (a\conc b \conc x \into X) $ described in Example~\ref{ex:multiterm}. If the function $f$ is defined as the canonical function $f^{can}_k$,  we get $\ov{t} \srewrites{2\cdot k} (a \conc b \conc b \into c) \conc (b \into c)$ and  $\ov{t} \srewrites{1\cdot k}  (b \conc b \into c) \conc (a \conc b \into c)$.
 \end{example}

\begin{remark}
Note that the canonical function $f^{can}$ has the same meaning of the
function  $\OO$ defined in \cite{BMMTT08}.
\end{remark}


  %
   %

\section{An application}\label{pho}

We used a prototype simulator for \short\ to run some simulations for the Pho regulation system described in Section~\ref{Sec:examples}.
In figure~\ref{stochmodelreg}) we use $\smapstoSug{k}$ to indicate that we use the canonical function
with kinetic constant k. The rates here are chosen manually following these considerations:
diffusion of phosphate across the outer membrane pores has the same speed in
both directions, PhoRP phosphatase activity is slow with respect to its kinase activity
in the unbound form, both these mechanism are slower than the simpler
binding and unbinding of phosphate to PhoR and the rules that describe translation
and traduction and protein degradation should be rather slow.

\begin{figure}[!h]
\begin{center}
\begin{equation}
\begin{aligned}
& Pi \conc (pore \conc x \into X) \smapstoSug{0.1} (pore \conc x \into Pi \conc X) \\
& (pore \conc x \into Pi \conc X) \smapstoSug{0.1} Pi \conc (pore \conc x \into X) \\
& Pi \conc (PhoR \conc x \into X) \smapstoSug{0.01} (PhoRP \conc x \into  X) \\
& (PhoRP \conc x \into  X) \smapstoSug{0.005} Pi \conc (PhoR \conc x \into X) \\
& (PhoR \conc x \into  PhoB \conc X) \smapstoSug{0.001} (PhoR \conc x \into PhoBP \conc X) \\
& PhoBP \conc PhoGenes \smapstoSug{0.0001} PhoBP \conc PhoGenes \conc PhoProt \\
& PhoProt \smapstoSug{0.00008} \emptyseq \\
& (PhoRP \conc x \into  PhoBP \conc X) \smapstoSug{0.0001} (PhoRP \conc x \into PhoB \conc X) \\ \nonumber
\end{aligned}
\end{equation}
\vspace{-0.8cm}
\caption{Stochastic rules for the phosphate regulation mechanism.}
\label{stochmodelreg}
\end{center}
\end{figure}

\begin{figure}[!h]
\begin{center}
\begin{minipage}{0.98\textwidth}
\begin{center}
\includegraphics[width=100mm]{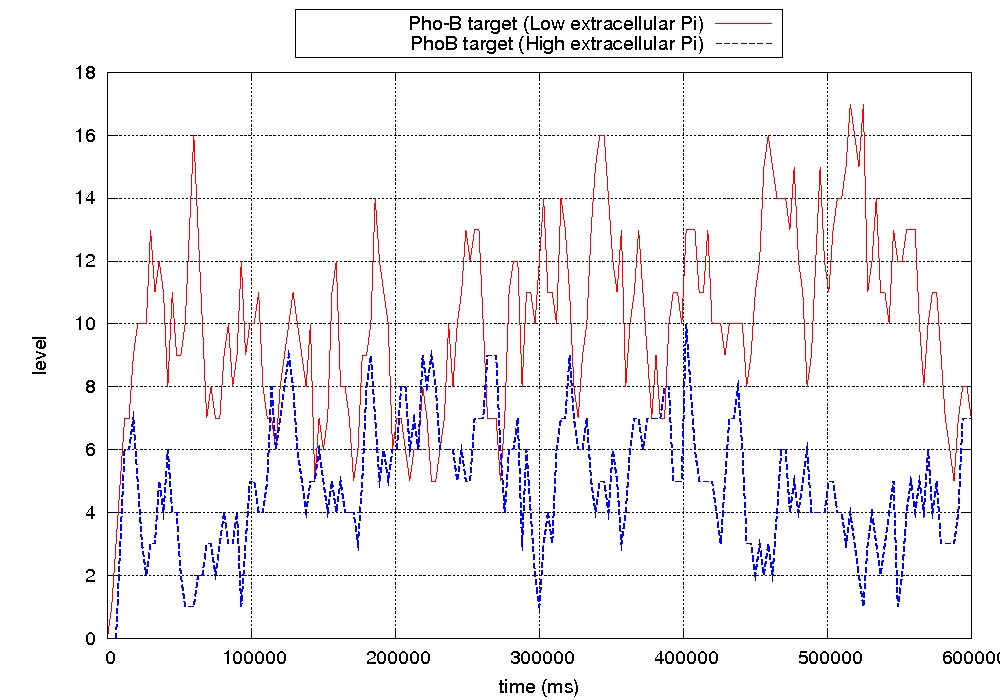}
\end{center}
\vspace{-0.5cm}
\caption{Levels of PhoB target proteins over time with different extracellular phosphate levels.}
\label{fig:20vs5}
\end{minipage}
\end{center}
\end{figure}

\begin{figure}[!h]
\begin{center}
\begin{minipage}{0.98\textwidth}
\begin{center}
\includegraphics[width=100mm]{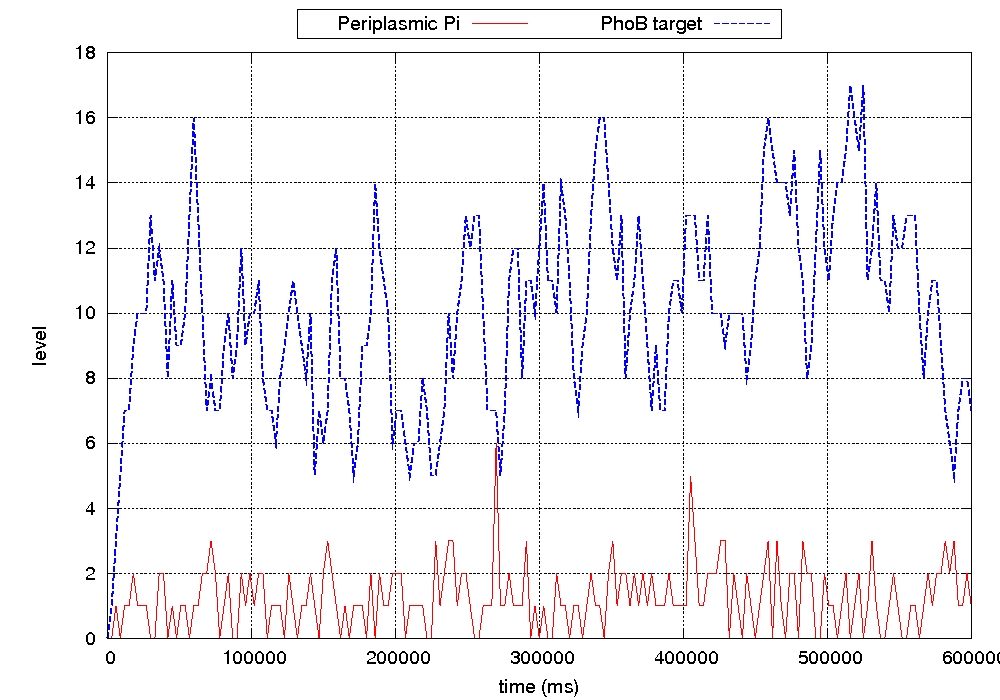}
\end{center}
\vspace{-0.5cm}
\caption{Levels of PhoB target proteins and periplasmic phosphate in the low extracellular phosphate conditions.}
\label{fig:Pivsprotein}
\end{minipage}
\end{center}
\end{figure}
\begin{figure}[!h]
\begin{center}
\begin{minipage}{0.98\textwidth}
\begin{center}
\includegraphics[width=100mm]{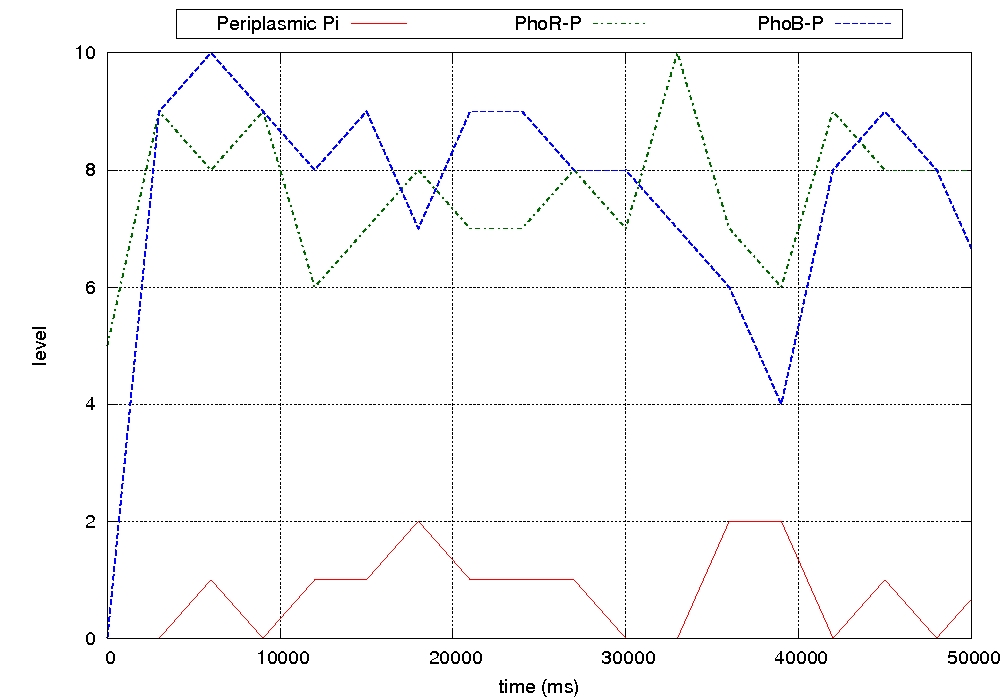}
\end{center}
\vspace{-0.5cm}
\caption{Levels of phosphorilated PhoB and phosphate bound PhoR in the low extracellular phosphate conditions.}
\label{fig:RPvsBP}
\end{minipage}
\end{center}
\end{figure}

We performed two simulations: both of them started with 10 PhoB, 5 PhoR, 5 PhoRP
and a PhoGenes element in a single bacterial cell. To reflect high and low
extracellular phosphate we started with 20 and 5 extracellular Pi, respectively.
In Figure~\ref{fig:20vs5} we show the comparison of the PhoProt levels obtained
in these simulations: as expected in the low phosphate condition the average
level of the PhoB target proteins is higher.
Figure~\ref{fig:Pivsprotein} shows the relation between PhoProt levels
and periplasmic phosphate for the low phosphate simulation:
oscillations in the protein levels are in sync with the levels of periplasmic phosphate: when the
latter increases the former decreases and viceversa.
Figure~\ref{fig:RPvsBP} shows the beginning of the same simulation but focuses on the relationships
between PhoRP, PhoPB and periplasmic phosphate: PhoRP levels follows with a little delay the phophate
ones, while PhoBP shows a clearly specular behaviour with respect to PhoRP, as expected.

In the future we will like to compare the results of these simulations with available experimental data~\cite{BKL07}.

\section{Conclusions}\label{conc}

As we have seen, \short\ allows to model cellular interaction, localisation and membrane structures.
Other formalisms were developed to describe membrane
systems. Among them we cite Brane Calculi~\cite{Car05} and
P-Systems~\cite{P02}.

\short\ can describe situations that cannot be easily captured by the
above mentioned formalisms, which consider membranes as atomic
objects. Representing the membrane structure as a multiset of the elements
of interest allows the definition of different functionalities
depending on the type and the number of elements on the membrane
itself.

From a qualitative point of view, our calculus is essentially a variant of CLS~\cite{BMMT06},
while, from a quantitative point of view, the possibility to model the speed of reactions using functions instead of constant rates makes the \short\ stochastic semantics more general with respect to the one of SCLS~\cite{BMMTT08}. In this sense, \short\ stochastic semantics is more similar to the one in~\cite{DGT09b}.

The current simulator for \short\ is implemented in OCaml (see~\cite{SCWC_SIM}). Since we do not need to analyse sequence structures and variables, the pattern matching algorithm has been simplified with respect to the simulator for SCLS described in~\cite{Sca07}.

We are also considering to enrich our analysis framework by taking into account statistical
model checking as done in~\cite{BCM09} for stochastic CLS models, and with techniques to approximate the value of unknown kinetics in the stile of~\cite{LMT04,LMT07,CGL09}.

\bibliographystyle{plain}
\bibliography{bio,fmb}

\end{document}